\documentclass[aps,prl,twocolumn]{revtex4}
\usepackage[dvips]{graphics}
\usepackage[dvips]{graphicx}
\usepackage{amsmath,amsthm,amsfonts,amssymb} 

\begin{document}
\title{Quantum Dynamics in the Thermodynamic Limit.}
\author{Jasper van Wezel}
\affiliation{
Theory of Condensed Matter, Cavendish Laboratory, University of Cambridge, Madingley Road, Cambridge CB3 0HE, UK
}
\date{\today}

\begin{abstract}
\noindent
The description of spontaneous symmetry breaking that underlies the connection between classically ordered objects in the thermodynamic limit and their individual quantum mechanical building blocks is one of the cornerstones of modern condensed matter theory and has found applications in many different areas of physics. The theory of spontaneous symmetry breaking however, is inherently an {\it equilibrium} theory, which does not address the {\it dynamics} of quantum systems in the thermodynamic limit.
Here, we will use the example of a particular antiferromagnetic model system to show that the presence of a so-called thin spectrum of collective excitations with vanishing energy --one of the well-known characteristic properties shared by all symmetry-breaking objects-- can allow these objects to also spontaneously break time-translation symmetry in the thermodynamic limit. As a result, that limit is found to be able, not only to reduce quantum mechanical equilibrium averages to their classical counterparts, but also to turn individual-state quantum dynamics into classical physics.
In the process, we find that the dynamical description of spontaneous symmetry breaking can also be used to shed some light on the possible origins of Born's rule.

\noindent
We conclude by describing an experiment on a condensate of exciton polaritons which could potentially be used to experimentally test the proposed mechanism.
\end{abstract}

\maketitle

\subsection{1: Introduction} \noindent
Combining many elementary particles into a single interacting system may result in collective behaviour that qualitatively differs from the properties allowed by the physical theory governing the individual building blocks. This realisation --immortalised by P.W. Anderson in his famous phrase 'More is Different'~\cite{Anderson72}-- not only forms the basis of much of the research being done in condensed matter physics today, but has also found applications in areas ranging from string theory to cosmology. The theory of Spontaneous Symmetry Breaking which formalises these ideas first took shape over fifty years ago~\cite{Landau37,Goldstone62,Anderson63:book,Anderson52,Anderson58,Nambu60}, and was completed in the context of quantum magnetism only two decades ago by the detailed description of the classical state as a combination of thin spectrum states, emerging as $N \to \infty$ because of the singular nature of the thermodynamic limit~\cite{Lieb62,Kaiser89,Kaplan90}. The same description of the classical state emerging from the thin spectrum has since been shown to also directly apply to the cases of quantum crystals, antiferromagnets, Bose-Einstein condensates and superconductors~\cite{vanWezel05,vanWezel06,vanWezel07,vanWezel07:SC,Birol07}.

The connection between the quantum mechanical properties of microscopic particles and the classical behaviour of symmetry broken macroscopic objects has now again come to the forefront of modern science because of our technological capability to create ever larger and heavier quantum superpositions in the laboratory. Superconducting flux qubits harbour counterrotating streams of supercurrent consisting of up to $10^{11}$ Cooper pairs~\cite{Wal00,Chiorescu00,Mooij:flux03}, while Bose Einstein condensates of the order of $10^5$ Rubidium atoms can be routinely brought into superpositions of different momentum states~\cite{Anderson95,Davis95,Stenger99,Kozuma99}; Young's double slit experiment has now been done using $C_{60}$ molecules instead of single photons or electrons~\cite{Zeilinger99}; and an experiment has even been proposed to create a Schr\"odinger cat-like state of a mesoscopic mirror superposed over a macroscopically discernible distance~\cite{Marshall03}.

Almost all of these experiments employ the rigidity associated with a spontaneously broken symmetry to create and manipulate their 'macroscopic' superpositions. Roughly speaking, the typical setup consists of a well defined, symmetry broken object in isolation (a superconductor, Bose Einstein condensate or crystal) which is brought into superposition by coupling it to a carefully selected quantum state. Although the theory of spontaneous symmetry breaking can be used to understand the stability and rigidity of macroscopic classical states such as superconductors or crystals, it says nothing about the quantum dynamics of such objects interacting with microscopic quantum states. The reason is that the standard description of spontaneous symmetry breaking is an inherently {\it equilibrium} description: it explains how macroscopic operators (such as the order parameter) can acquire finite expectation values and still be in stable equilibrium, but it does not say anything about the {\it dynamics} of these objects away from equilibrium.

A theoretical framework which does addresses the interaction of a macroscopic object with its microscopic quantum mechanical environment, is the study of decoherence~\cite{Zurek81,Joos85,CaldeiraLeggett}.  The basic idea of decoherence is that the entanglement  of a certain quantum state with the many states of its environment can lead that state to behave effectively classically as long as the environmental states remain unobservable. This phenomenon has many practical implications, not in the least in the field of quantum information technology, where decoherence forms the main hurdle to be overcome in the race towards a working quantum computer. In the description of the interaction of a single macroscopic object with a single quantum state however, the theory of decoherence cannot be applied. The problem is that decoherence has to always refer to the properties of an {\it ensemble average}: after deciding which of the environmental degrees of freedom cannot be measured, one has to trace them out of the full density matrix describing the combined system of object and environment. Doing this (partial) trace is exactly equivalent to taking the quantum mechanical expectation value of the operators describing the unobserved states, and as such is only defined within an ensemble and cannot be used to say anything about the outcomes of {\it single-shot} experiments~\cite{Adler,Bassi}. 

In this paper, we will develop a description of dynamical spontaneous symmetry breaking that is meant to augment the earlier theories of equilibrium spontaneous symmetry breaking and decoherence in the areas where these theories do not apply. It will describe the quantum dynamics of individual experiments in which macroscopic and microscopic systems are allowed to interact. We will find that the presence of thin spectrum states in symmetry-broken objects allows these systems to also spontaneously break the unitarity of quantum mechanical time evolution. This result explains why truly macroscopic objects do not dynamically delocalise even if they are allowed to interact and entangle with an observable quantum mechanical environment. At the same time it also sheds light on what happens if the classical state is forced into a superposition state by an interaction with a carefully chosen quantum state.

In section 2 we start out with a short review of the equilibrium theory of spontaneous symmetry breaking. The role of the thin spectrum and the singular nature of the thermodynamic limit will be highlighted. In section 3 we then review the theory of decoherence and point out why it refers only to ensemble averages. We then turn to dynamic spontaneous symmetry breaking in section 4, using a model antiferromagnetic system as an example. It is argued there that the thin spectrum states and the thermodynamic limit can cooperate to allow the spontaneous breakdown of quantum mechanical unitarity. The resulting dynamics of a single quantum state in the thermodynamic limit is studied. We then continue in section 5 by describing the fate of a macroscopic object that is forced into a quantum superposition through the interaction with a microscopic quantum state. The results are again clarified using the example of the model antiferromagnet, and are shown to shed new light on the emergence of Born's rule. Finally, in section 6, we describe a possible experimental test of the ideas of sections 4 and 5 using a condensate of exciton polaritons. We end in section 7 with a summary and conclusions.

\subsection{2: Equilibrium Spontaneous Symmetry Breaking} \noindent
Classically, spontaneous symmetry breaking just corresponds to the evolution from a high symmetry metastable state into a ground state with lower symmetry. 
%
Quantum mechanically however, the situation becomes a bit more involved. 
%
%
First of all, there are in general nonzero tunneling matrix elements between different symmetry broken states, so that strictly speaking time evolution should cause any symmetry broken state to spread out and restore its symmetry.
In practise though, this finite lifetime of a symmetry broken state can be easily shown to be long compared to the age of the universe for any realistic macroscopic system. Secondly, the symmetry broken states of a finite size system do not have to be ground states. In fact, they usually are not even eigenstates of the system. 
%
%

To establish how the system can end up in a state that is not an eigenstate of the underlying Hamiltonian, we will here use the specific example of the Lieb-Mattis model~\cite{Lieb62,Kaiser89,Kaplan90,vanWezel06,vanWezel07}. This model is defined by the Hamiltonian:
\begin{align}
H_{\text{LM}} 
&= \frac{2 J}{N} {\bf S}_A \cdot {\bf S}_B \nonumber \\
&= \frac{J}{N} \left[ S^2 - S_A^2 - S_B^2 \right].
\label{Hlm}
\end{align}
Here $N$ spin-$\frac{1}{2}$s are distributed over a bipartite lattice, with ${\bf S}_{A/B}$ the total spin of the $A/B$ sublattice and $S^z_{A/B}$ its $z$-projection. Each spin on the $A$ sublattice thus has an interaction with every spin on the $B$ sublattice and vice versa. The positive interaction strength $J$ is divided by $N$ to make the model extensive. $S$ is the total spin of the combined sublattices: $S=S_A+S_B$.
The reason for considering specifically the Lieb-Mattis model with its infinitely long ranged interactions, is that it captures the relevant physics of a broad class of Heisenberg models with short ranged interactions. To say that a particular model for an antiferromagnet is invariant under SU(2) spin rotations is equivalent to stating that its Hamiltonian commutes with the total spin operator: $\left[H,S^2 \right]=0$. It is thus immediately obvious that total spin is a good quantum number for any isotropic antiferromagnet and that all their eigenstates can be labelled by such a total spin quantum number. For the description of the collective properties of the system as a whole (i.e. strictly infinite wavelength), the total spin is the only relevant part of the Hamiltonian. As far as the total spin is concerned, the Lieb-Mattis model coincides exactly with all other antiferromagnetic models. That is to say, if one takes {\it any} model for an antiferromagnet with short ranged interactions (such as for example the nearest neighbour Heisenberg model) and looks at the model in Fourier space, then the $k=0$ and $k=\pi$ modes together form {\it exactly} the Lieb Mattis-Hamiltonian~\cite{vanWezel06,vanWezel07}. At the same time, the finite wavelength, $k\neq 0,\pi$ modes are gapped and dispersionless in the Lieb-Mattis model due to the infinite ranged interactions, which makes it ideally suited for studying just the collective behaviour of antiferromagnets. The discussion of this model can also be easily adapted to describe spontaneous symmetry breaking in quantum crystals, superconductors and Bose-Einstein condensates~\cite{vanWezel06,vanWezel07,vanWezel07:SC,Birol07}.

From the second expression in equation \eqref{Hlm} it is immediately clear that the ground state of the Lieb-Mattis system is a singlet state with zero total spin. This non-degenerate ground state is isotropic in spin space and thus fully respects the symmetry of its Hamiltonian. The heart of the workings of spontaneous symmetry breaking lies in the realisation that every many-particle Hamiltonian which possesses a continuous symmetry that is unbroken in its ground state (such as the Lieb-Mattis Hamiltonian), gives rise to a tower of low-energy states called the \emph{thin spectrum}~\cite{vanWezel05,vanWezel07}. The states in this thin spectrum represent global (infinite wavelength) excitations that can be seen as the centre of mass properties of the collective system~\cite{vanWezel07,Birol07}. In the present model of equation~\eqref{Hlm} the thin spectrum consists of total spin states, which only cost an energy of order $J/N$ to excite. These states thus become degenerate with the symmetric ground state in the thermodynamic limit. They are called the {\it thin} spectrum of the model because of the vanishing weight that these states have in the partition function.  Excitations that change the size of the sublattice spins are separated from the ground state by an energy gap of size $J$, and can thus be ignored in the present (low energy) discussion. Without loss of generality we also set the $z$ projection of the total spin to be zero from here on.

The crucial observation is now that the strength of the field needed to give rise to a fully ordered ground state depends on the total number of particles, $N$, in the system. Because the energy separation between two consecutive thin spectrum states scales as $1/N$, the field strength necessary to explicitly break the symmetry decreases with system size. In the thermodynamic limit (where $N \to \infty$) all of the thin spectrum states collapse onto the ground state to form a degenerate continuum of states. Within this continuum even an \emph{infinitesimally} small symmetry breaking field is enough stabilise a fully ordered, symmetry broken ground state. The system is thus said to spontaneously break its symmetry in that limit. To make this explicit in the present model, we add a symmetry breaking staggered magnetic field to the Hamiltonian:
\begin{align}
H_{\text{LM}}=\frac{2 J}{N} {\bf S}_A \cdot {\bf S}_B - B \left( S^z_A - S^z_B \right).
\label{HlmSB}
\end{align}
The staggered magnetisation only has non-zero matrix elements between consecutive thin spectrum levels~\cite{vanWezel07}:
\begin{align}
\left< S' \left| S^z_A - S^z_B \right| S \right> & = \delta_{S'+1,S} f_S  + \delta_{S'-1,S} f_{S'} \nonumber \\
& \simeq \frac{N}{4} \left( \delta_{S'+1,S}  + \delta_{S'-1,S} \right),
\label{Bmatrix}
\end{align}
where $f_S \equiv \sqrt{\{ S^2 [ \left(S_A+S_B+1\right)^2 -S^2 ] \} /\{4S^2-1\}}$, and the approximation in the last line holds if $S_A=S_B=N/4$ and $1 \ll S \ll N$~\cite{Kaiser89}. The Schr\"odinger equation for the Lieb-Mattis model, $H_{\text{LM}} |n\rangle = E^n |n\rangle$, can be expanded in the total spin basis using $|n\rangle \equiv \sum_S u_S^n |S\rangle$. Upon taking the continuum limit it then reads
\begin{align}
-\frac{1}{2}\frac{\partial^2}{\partial S^2} u_S^n + \frac{1}{2} \omega^2 S^2 u_S^n = \nu^n u_S^n,
\label{un}
\end{align}
with $\omega=2/N \sqrt{J/B}$ and $\nu^n = 2 E^n / (B N) + 1$. This equation describes a harmonic oscillator and its eigenfunctions are given in terms of the well known Hermite polynomials. The expansion of these harmonic wavefuntions in the total spin basis brings to the fore the crucial role played by the thin spectrum in the mechanism of spontaneous symmetry breaking: because the total spin states all become degenerate in the limit $N \to \infty$, it then becomes arbitrarily easy to create the antiferromagnetic N\'eel state $|n=0\rangle = \sum_S u_S^0 |S\rangle$. Mathematically this translates into the non-commuting limits for the equilibrium expectation values of the order parameter~\cite{vanWezel07}
\begin{align}
\lim_{N \to \infty} \lim_{B \to 0} \left<\frac{S_A^z-S_B^z}{N/2}\right> &= 0 \nonumber \\
\lim_{B \to 0} \lim_{N \to \infty} \left<\frac{S_A^z-S_B^z}{N/2}\right> &= 1.
\label{limits}
\end{align}
The same instability can also been seen by looking at the energy of the ground state in the presence of the symmetry breaking field. That energy is proportional to $-NB$ and thus an infinite amount of energy could be gained in the thermodynamic limit by aligning with an infinitesimally small symmetry breaking field.

An alternative, equivalent way of phrasing this singular property of the thermodynamic limit is to say that the limits of equation \eqref{limits} imply that even in the absence of $B$, quantum fluctuations of the order parameter which tend to disorder the symmetry broken state take an infinitely long time to have any measurable effect on a truly macroscopic system. Under equilibrium conditions, the system will thus be stable in a symmetry broken state that is not an eigenstate of its Hamiltonian. 

Strictly speaking equation~\eqref{limits} only allows truly infinite-size systems to spontaneously select a direction for their sublattice magnetisation. A large, but not infinitely large, system requires a finite symmetry breaking field to stabilise one of the symmetry broken states over the exact ground state. A true staggered magnetic field that points up on each site of the $A$ sublattice and down on the $B$ sublattice does not exist in nature. Because the strength of the required field becomes increasingly weaker as the size of the antiferromagnet grows, it can be argued however that {\it any} field which has a component that resembles a staggered magnetic field will be enough to stabilise the symmetry broken state in a large enough antiferromagnet. Such a weak staggered field could be provided in practise by magnetic impurities, local fields or even by a second antiferromagnet at an ever increasing distance from the first.

\subsection{3: Decoherence} \noindent
We have seen in the last section how spontaneous symmetry breaking enables a macroscopic collection of quantum mechanical particles to occur in an effectively classical symmetry broken state under equilibrium conditions. A different route from quantum mechanics to effectively classical behaviour is provided by the process of decoherence. Decoherence happens on all length scales (i.e. it does not require the object of interest to be macroscopic), and is a direct consequence of the inability of observers to monitor each and every degree of freedom of a typical quantum environment. At the heart, decoherence is the process in which a carefully prepared quantum state gets entangled with different states in its environment. Because the observer cannot measure all states of the environment, he can see only part of the final entangled state, and this partial state looks effectively classical. In this section we will use the Lieb-Mattis model as an example to highlight the different conceptual steps involved in the decoherence process.

Consider the Hamiltonian of equation~\eqref{Hlm}. Its eigenstates can be written as $|m,S\rangle \equiv |S_A=S_B=N/4-m/2,S\rangle$ (where we have assumed $S^z=0$ and $S_A=S_B$ without loss of generality). The excitations $m$ represent magnons or spin waves while the excitations $S$ form the thin spectrum of this model. Because the thin spectrum excitations make only a vanishingly small contribution to the free energy of the Lieb-Mattis antiferromagnet if $N$ is large, they will be very hard to observe experimentally (for relatively small $N$ the thin spectrum states of molecular antiferromagnets can and have been experimentally observed~\cite{Waldmann03,Waldmann05}). For large systems we can thus regard the thin spectrum as a 'quantum environment' for the magnon excitations. To study decoherence in this system we will first prepare a superposition state in the magnon sector, then we will let the magnon and the thin spectrum excitations interact and become entangled, and finally we will disregard the thin spectrum states and find that magnon states on their own have become an effectively classical mixture.

To prepare the initial magnon superposition, let us assume that we can access the exact ground state of the $N$-spin system and subsequently let it interact with a separate two-spin singlet state $\sqrt{1/2} \left[ |\uparrow \downarrow \rangle - |\downarrow \uparrow\rangle \right]$ through the instantaneous interaction defined by:
\begin{align}
H = \left\{ \begin{array}{ll}  
\frac{2 J}{N} {\bf S}_A \cdot {\bf S}_B + J {\bf S}_1 \cdot {\bf S}_2  & \text{for}~t<0 \\
\frac{2 J}{N+2} \left({\bf S}_A + {\bf S}_1 \right) \cdot \left( {\bf S}_B + {\bf S}_2 \right) & \text{for}~t>0.
\end{array} \right.
\label{Hint}
\end{align}
Here ${\bf S}_{1/2}$ refer to the two initially separated spins, and the interaction is turned on at time $t=0$. In terms of the eigenstates of the Hamiltonian at positive times, the initial state can easily be shown to correspond to the state $\sqrt{1/2} \left[|m=0,S=0\rangle - |m=2,S=0\rangle \right]$ for large $N$ (where now $m$ and $S$ refer to the $N+2$-spin system). That is, for large $N$ the initial state of the two-spin system is encoded in the number and relative phase of the magnon excitations in the final state~\cite{masterthesis}.

Next, we would like to entangle the magnons with the thin spectrum so that the quantum information initially encoded in the magnon states is spread out over the environment. One way of achieving this is to instantaneously introduce a symmetry breaking field $B \left( S^z_A + S_1^z - S^z_B - S_2^z \right)$ into the Hamiltonian at some positive time $t_0$. After some straightforward algebra the state of our systems at times $\tau=t-t_0$ is then found to be 
\begin{align}
\hspace{-6pt} \left| \psi \right> = \sqrt{\frac{1}{2}} \sum_{n,S} u_S^n u_0^n \left[ e^{-\frac{i}{\hbar}E^n_0 \tau} \left| 0,S \right> - e^{-\frac{i}{\hbar}E^n_2 \tau} \left| 2,S \right> \right]
\label{state}
\end{align} 
where $u_S^n$ are the harmonic wavefunctions defined in equation~\eqref{un} and $E^n_m$ is the energy of the $n^{\text{\tiny{th}}}$ harmonic wavefunction in the presence of $m$ magnons. We can write this final entangled state in the form of a density matrix through the definition $\rho(\tau) = | \psi \rangle \langle \psi |$.
Notice that all the quantum information encoded in the initial two-spin singlet state is still present in the final density matrix $\rho(\tau)$. Because purely quantum mechanical time evolution is always strictly unitary, time inversion symmetry is automatically preserved and there is always a way (at least in principle) to evolve the system back to its original state. If we now decide that the thin spectrum states are unobservable, and trace them out of our density matrix~\cite{Neumann55}, we end up with a reduced density matrix describing the dynamics of the magnons only. In doing so however, the time inversion symmetry is lost along with some of the quantum information. To be specific, the reduced density matrix $\rho_{\text{red}}$ will be given by: 
\begin{align}
\rho_{\text{red}}(\tau) &= \text{Tr}_{\text{thin}} \phantom{.} \rho(\tau)  \nonumber \\
&= \sum_{S} \left<S \left| \right. \psi \right> \left< \psi \left| \right. S\right>  \nonumber \\
&= \sum_{m,m'}  \left|m\right> \left\{ \sum_S \psi^{\phantom{*}}_{m,S} \psi^{*}_{m',S} \right\} \left<m'\right|.
\label{reduced}
\end{align}
In the last line we have written the entangled wavefunction as $|\psi\rangle=\sum_{m,S} \psi(m,S) |m\rangle|S\rangle$ to show explicitly that taking the partial trace over the thin spectrum states is equivalent to calculating the usual quantum mechanical ensemble-averaged expectation value with respect to these states. 

To complete the analysis of our model interaction, we should explicitly calculate the reduced density matrix elements of equation~\eqref{reduced}. The diagonal elements of the resulting $2$x$2$ matrix are easily seen to be $1/2$. For the off-diagonal elements the calculation involves a summation over terms which differ only by the phase factor $e^{-\frac{i}{\hbar}(E^n_0-E^n_2) \tau}$. After some algebra one finds that these phases interfere destructively~\cite{masterthesis}, so that after a time $\tau_{\text{coh}}\sim \hbar / \sqrt{J B}$ the reduced density matrix becomes effectively diagonal. We thus find that the initial, pure density matrix loses its coherence and becomes a diagonal, mixed reduced density matrix within a time $\tau_{\text{coh}}$. Because for large enough $N$ the environmental states are unobservable this constitutes a 'for all practical purposes' reduction from quantum to classical physics within the ensemble average. In any one single, individual experimental realisation of the above procedure however, one ends up with the full density matrix defined by equation~\eqref{state}, and one cannot use the expectation values of equation~\eqref{reduced} to conclude anything about that one specific experiment. In particular, in the classic Young's double slit experiment, the observation that each single electron produces only a single dot on the photographic plate, can {\it not} be explained by invoking decoherence and averaging over the many degrees of freedom of the plate~\cite{Tonomura89,Adler,Bassi}.

Although the presence of the thin spectrum can lead to decoherence in real qubits~\cite{vanWezel05}, the interaction of the Lieb-Mattis antiferromagnet and the two-spin state considered in this section is of course a highly pathological example. In reality there will never be infinite ranged interactions, instantaneous changes to the Hamiltonian or full experimental control over the prepared states. Moreover, experiments typically involve finite temperatures and external environments that do not resemble the thin spectrum states of our model. However, the general idea of constructing a meaningful quantum superposition, letting it interact and entangle with its environment, and then looking only at the result averaged over the environmental degrees of freedom to find decoherence, remains essentially unaltered in more realistic situations~\cite{CaldeiraLeggett}. In particular the conclusion that the the theory of decoherence is applicable only within the realm of ensemble averages remains intact throughout.

\subsection{4: Dynamic Spontaneous Symmetry Breaking} \noindent
As we have seen, both the theory of spontaneous symmetry breaking and the theory of decoherence have only a limited domain of applicability. Because macroscopic states typically have a lot of interaction with their environments, decoherence explains the reduction of pure macroscopic states to mixed states in situations where not all degrees of freedom can be explicitly monitored, but only in an (ensemble) averaged sense. Spontaneous symmetry breaking on the other hand can be used to demonstrate the stability of macroscopically ordered, classical states using the singular nature of the thermodynamic limit and the properties of the thin spectrum, but only under equilibrium conditions. The most general situation involving macroscopic objects --that of individual-state quantum dynamics in the thermodynamic limit-- cannot be addressed within either of these frameworks.

In this section we will show that the presence of a thin spectrum in objects that can undergo spontaneous symmetry breaking also allows these objects to spontaneously break the (unitary) time translation symmetry of quantum mechanical time evolution. The resulting dynamical version of the process of spontaneous symmetry breaking naturally leads to the observed stability of macroscopic objects even in the presence of interactions with a quantum environment.

The approach to spontaneously breaking time translation symmetry is exactly analogous to the spontaneous breaking of more usual symmetries: we will introduce a vanishingly small non-unitary perturbation to the free Hamiltonian and demonstrate that this results in a qualitative change to the dynamics of a macroscopic object, even in the limit of taking the field strength to zero. The conclusion must thus be that the quantum dynamics of these macroscopic objects is infinitely sensitive to any non-unitary perturbation of the type considered. In other words: purely unitary quantum dynamics is unstable in the thermodynamic limit in the same way that the total spin singlet state of a macroscopic antiferromagnet is an unstable state under equilibrium conditions. As a result the unitary time translation symmetry of macroscopic quantum objects will be spontaneously broken and give rise instead to classical dynamics.

At this point one may wonder about the physical origin of the symmetry breaking field. As with the usual equilibrium symmetry breaking, large but finite sized systems will require a very small but nonetheless finite symmetry breaking field. Non-unitary fields however are strictly forbidden in quantum theory. The origin of a non-unitary symmetry breaking field must therefore lie outside of quantum mechanics. There are many possible candidates that could in principle insert a vanishingly small non-unitary correction into quantum mechanics. A notable example is the theory of general relativity, in which general covariance rather than unitarity is the guiding principle. Because of this, gravity has (in a different setting) been considered before as a possible non-unitary influence on mesoscopic systems~\cite{Diosi89,Penrose:96,vanWezel:penrose}. In this paper we will not speculate about the possible origins of the non-unitary field, but merely recognise that there are non-unitary physical theories outside of the realm of quantum mechanics, and that only an infinitesimally small contribution from one of these sources would be enough to spontaneously break the unitarity of quantum dynamics in the thermodynamic limit.

We thus consider once again the Lieb-Mattis model for an antiferromagnet, but now in the presence of a non-unitary symmetry breaking field:
\begin{align}
H=\frac{2 J}{N} {\bf S}_A \cdot {\bf S}_B + i b \left( S^z_A - S^z_B \right).
\label{Hdssb}
\end{align}
The rationale of which specific form of non-unitary field is to be included in this equation is again exactly analogous to the case of equilibrium spontaneous symmetry breaking: one should in principle consider every conceivable field. The system will of course be stable with respect to the vast majority of them, but as long as there is one that has an effect in the limit in which its strength is sent to zero, the system will be unstable. In the equilibrium case considered before, we have seen that the symmetric singlet state is unstable with respect to a staggered magnetic field along the $z$-axis. We could have also considered other symmetry breaking fields, such as a uniform magnetic field along the $z$-axis. It is easy to show however that such a field would not lead to the non-commuting limits of equation \eqref{limits}. The Lieb-Mattis system is thus shown to be unstable under equilibrium conditions with respect to antiferromagnetic ordering, but not with respect to ferromagnetic ordering. The situation in the dynamical case is analogous: most fields have no effect on the quantum dynamics of the system if their strength is sent to zero; But as soon as there is one field that does influence the dynamics even if it is infinitesimally weak, the dynamics is found to be unstable. Notice also that in the equilibrium case, the antiferromagnet is in fact unstable towards staggered magnetic fields along any axis. In practise, the resulting orientation of the order parameter is therefore randomly chosen, just as in the case of classical symmetry breaking. In equation ~\eqref{Hdssb} we have chosen a non-unitary version of the staggered magnetic field to break time translational symmetry, because to be able to have an effect in the thermodynamic limit, the symmetry breaking field must couple to the order parameter of the system. The orientation along the $z$-axis rather than any other axis is chosen for convenience only.

The time evolution operator $U(t) \equiv \exp (-iHt/\hbar)$ implied by equation~\eqref{Hdssb} has a non-unitary component, and thus no longer automatically conserves the total energy of the system (defined as $\langle H \rangle$ with $b \to 0$). This problem is automatically solved in the thermodynamic limit though. The staggered magnetisation only has non-zero matrix elements between consecutive states in the thin spectrum (see equation~\eqref{Bmatrix}). Since all thin spectrum states become degenerate with the ground state in the limit $N \to \infty$, the time evolution defined through $H$ cannot alter the total energy of the system in that limit. Other problems that are usually associated with non-unitary quantum dynamics (conservation of normalisability, commutativity, and so on) are likewise automatically solved in the limit of vanishing $b$ and large $N$.

To visualise the time evolution defined by $U(t)$, consider a general initial state $| \psi(t=0) \rangle = \sum_S \psi_S(t=0) |S\rangle$ (we again take $S_A$ and $S_B$ maximal and $S^z=0$). Using $| \psi(t) \rangle = U(t) | \psi(0) \rangle$ we then find the generalised (non-unitary) Schr\"odinger equation to be:
\begin{align}
\hspace{-6pt} \dot{\psi}_S = \frac{-i}{\hbar}\frac{J}{N} S(S+1) \psi_S + \frac{b}{\hbar} \left( f_{S+1} \psi_{S+1} + f_S \psi_{S-1} \right)
\label{psidot}
\end{align}
with $f_S$ the matrix elements defined in equation~\eqref{Bmatrix}. This differential equation for the time evolution of a general initial wavefunction cannot easily be solved analytically (taking the limit in which $S$ becomes a continuous variable and $1 \ll S \ll N$, there is a solution in terms of Whittaker functions, but this explicit solution is not very enlightening for our present purposes). One can however integrate equation~\eqref{psidot} forward in time numerically, and we can study the effect of the unitarity breaking field by comparing the resulting time evolutions of different initial states. Two initial states of particular interest are the completely symmetric singlet state and the symmetry broken antiferromagnetic N\'eel state. 

\begin{figure}[t]
\center
\includegraphics[width=\columnwidth]{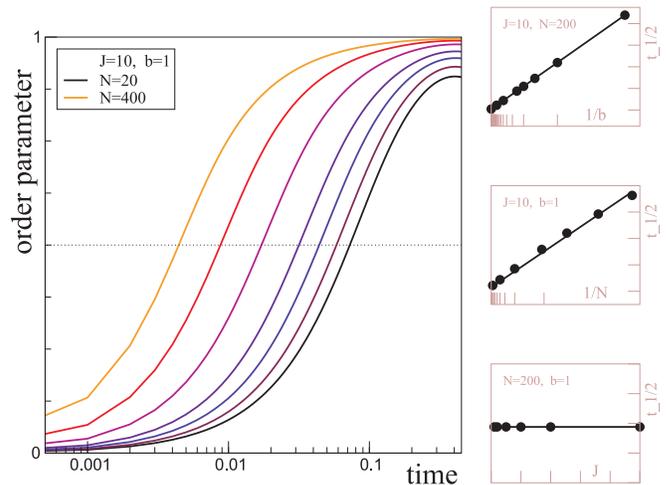}
\caption{(Color online) Left: The staggered magnetisation as a function of time. To make the plot the values $J=10$ and $b=1$ were used, and time was measured in units of $\hbar s$. The curves range from $N=20$ (rightmost curve) to $N=400$ (leftmost curve) and represent the evolution starting from the completely symmetric singlet state. \\ Right: The dependence of the halftime on the parameters of the model. The top plot shows that $t_{1/2} \propto 1/b$, the middle plot that $t_{1/2} \propto 1/N$, and the bottom plot that $t_{1/2}$ is independent of $J$.}
\label{symplot}
\end{figure}
In the case of the symmetric initial state the time evolution of equation~\eqref{psidot} leads the unitarity breaking field to amplify the weight of states with a finite order parameter (i.e. its component in the wavefunction becomes a monotonously increasing exponential function), so that a fully ordered state is quickly formed. In figure~\ref{symplot} the time evolution of the order parameter is shown for different values of $b$, $J$ and $N$. It is immediately clear that the half-time associated with the reduction towards an ordered state must be proportional to $1/(Nb)$, so that the thermodynamic limit in this case is found to be a singular limit: if we let $b$ go to zero before sending $N$ to infinity, the symmetric singlet state remains an eigenstate of $H$ and under time evolution it can only pick up a total phase; if on the other hand even just an infinitesimally small field $b$ is present while the thermodynamic limit is taken, the time evolution governed by $H$ gives rise to an instantaneous reduction of the symmetric state to the fully ordered state with the order parameter pointing in the direction of $b$. Analogous to the equilibrium description, this non-commuting order of limits signals the sensitivity of the system to even infinitesimally small perturbations. In this case it is the unitary time translational symmetry of quantum dynamics itself that is spontaneously broken, and as a result the symmetric singlet state will be spontaneously and instantaneously reduced to an ordered N\'eel state.

\begin{figure}[t]
\center
\includegraphics[width=\columnwidth]{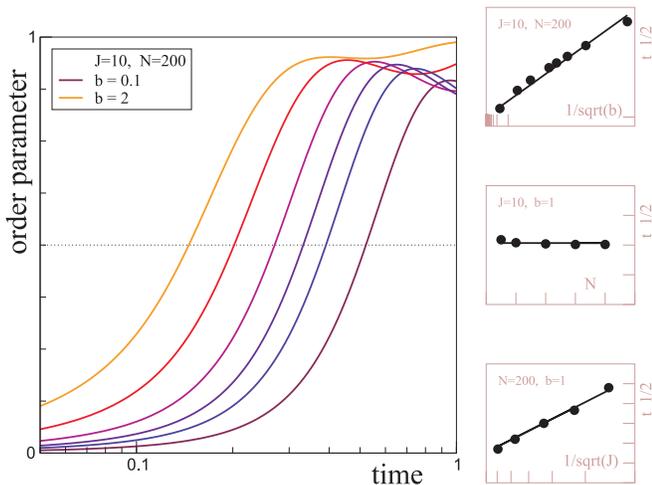}
\caption{(Color online) Left: The staggered magnetisation along the $z$ axis as a function of time. To make the plot the values $J=10$ and $N=200$ were used, and time was measured in units of $\hbar s$. The curves range from $b=0.1$ (rightmost curve) to $b=2$ (leftmost curve) and represent the evolution starting from the state with full antiferromagnetic order along the $x$ axis. \\ Right: The dependence of the halftime on the parameters of the model. The top plot shows that $t_{1/2} \propto \sqrt{1/b}$, the middle plot that $t_{1/2}$ is independent of $N$, and the bottom plot that $t_{1/2} \propto \sqrt{1/J}$.}
\label{xtozplot}
\end{figure}
Starting from a fully ordered initial state, the picture changes drastically. The state which has antiferromagnetic order aligned with the field $b$ to start with, will not be influenced at all. That state is just a stable state with respect to the generator of time evolution $U(t)$. The evolution of the initial state with full N\'eel order along the $x$ axis (at a $90$ degree angle with the field $b$) is shown in figure~\ref{xtozplot}. The effect of the presence of the unitarity breaking term is clearly to align the initial order parameter with the field $b$. The timescale on which this process takes place however is proportional to $\sqrt{1/(Jb)}$. This time is just the ergodic time of the Lieb-Mattis system and it becomes infinitely long in the thermodynamic limit with a vanishing symmetry breaking field. The difference between this 'turning time' and the 'ordering time' of the symmetric state considered before is due to the fact that for large objects any fully ordered state becomes exactly orthogonal to all differently ordered states, while the symmetric state always keeps a finite overlap with all of them~\cite{Anderson:SolidState}. The lifetime of the symmetric state is therefore determined simply by the strength of the amplification due to the unitarity breaking field,  while the turning time of a fully ordered initial state is set by the ergodic time of the system. Starting from the ordered state, the limit $N \to \infty$ is thus no longer singular: regardless of the size of $N$, the limit $b \to 0$ will reduce any dynamics to just the standard quantum mechanical time evolution. The dynamics of the ordered state, in other words, is stable with respect to the unitarity breaking field $b$. 

Summarising, it has become clear that even an infinitesimally small unitarity breaking field is enough in the thermodynamic limit to instantaneously convert a symmetric initial state into a fully ordered state. Once such an ordered state has been formed however, it is stable with respect to any differently aligned unitarity breaking field. The former instability explains why the interaction with its environment cannot cause the wavefunction of a macroscopically ordered state to spread. After all, the more symmetric, spread-out wavepacket would be an unstable state, and it would spontaneously and instantaneously be brought back to the ordered state. At the same time the stability of the macroscopically ordered state itself ensures that such a state cannot spontaneously change the direction of its order parameter.

In the above analyses we have only considered symmetry breaking fields that are constant in time. Because of the dynamical nature of the spontaneous symmetry breaking process, it would actually be more natural to also include time dependent non-unitary fields. Since the strength of the field is taken to be infinitesimal, the time dependence of such a field must lie in its spatial orientation. As we have seen, ordered states are stable with respect to any orientation of the symmetry breaking field, and will thus also be stable with respect to a fluctuating field. The symmetric state on the other hand is sensitive to the direction of the field $b$: it is along this direction that the ordered state is formed. A fluctuating symmetry breaking field will thus cause the quantum dynamics of a symmetric state to amplify different orientations of the order parameter at different times. As a result both the direction and the size of the overall staggered magnetisation will undergo a random walk. As soon as the size of the magnetisation is large enough however, the dynamics again reduces to that of the ordered state, and the influence of the symmetry breaking field will no longer be felt. Because the symmetric state reacts infinitely fast to an infinitesimal perturbation in the thermodynamic limit, the whole process of undergoing a random walk and picking out an orientation for the order parameter will still be effectively instantaneous, and the earlier conclusions about the stability of quantum dynamics in the thermodynamic limit remain unaltered even in the presence of a fluctuating field.

\subsection{5: Macroscopic Superpositions and Born's Rule} 
\begin{figure}[tb]
\center
\includegraphics[width=0.8\columnwidth]{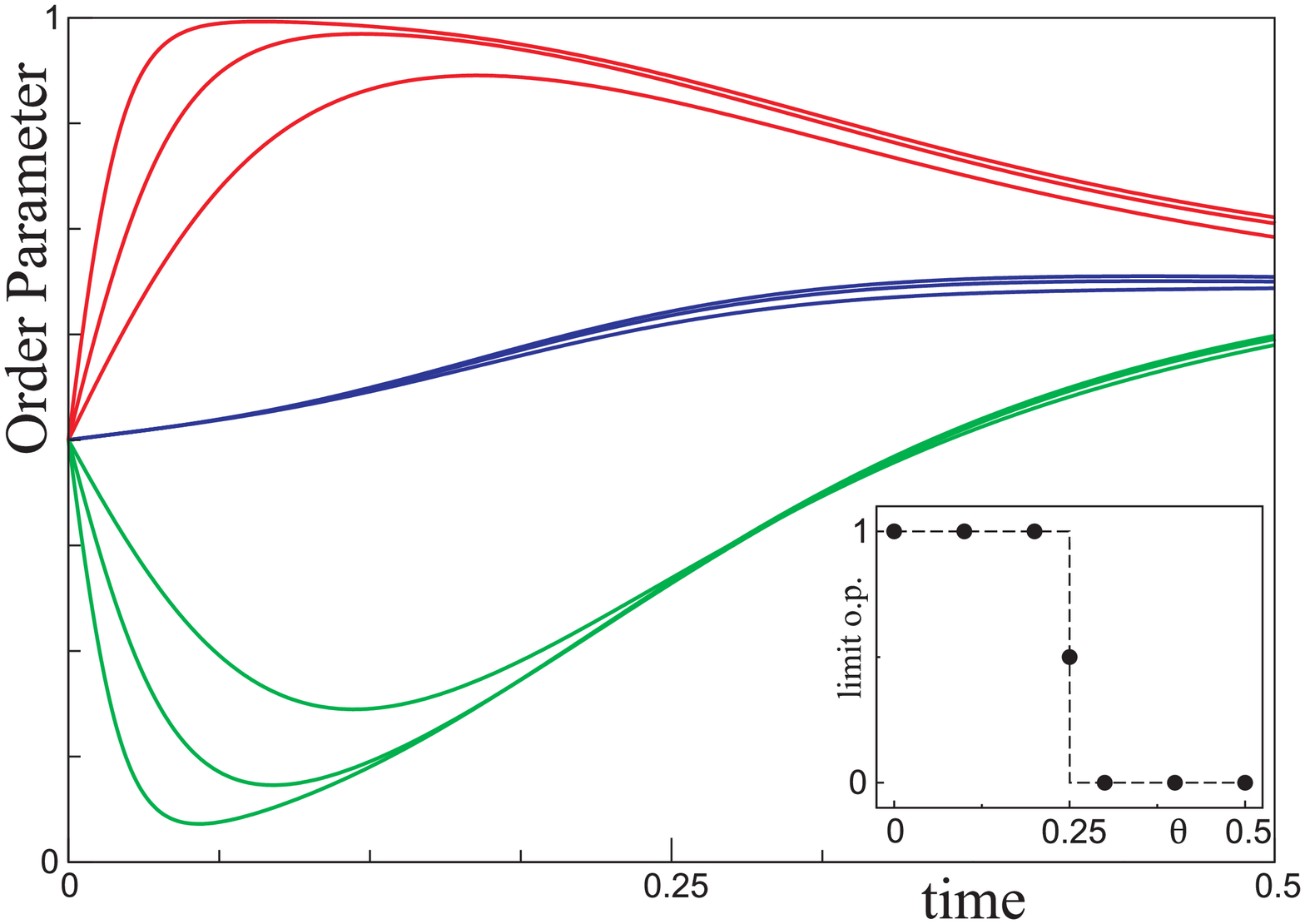}
\caption{(Color online) The evolution of the order parameter as a function of time (in units of $\hbar s$) for different constant orientations of the unitarity breaking field. Each set of three curves consists of different numbers of spins which are initially prepared in an equal-weight superposition of being ordered along the $z$ axis and along the $x$ axis. The angle $\theta$ between the unitarity breaking field and the $z$ axis $0.2~\pi$ for the upper set, $0.25~\pi$ in the middle and $0.3~\pi$ for the lowest set. The inset shows the fate of the order parameter in the thermodynamic limit, as a function of $\theta$.}
\label{theta}
\end{figure}

\noindent
Having established that a macroscopically ordered state is stable and will not be driven into a quantum superposition of differently ordered states by its environment, the question arises what would happen to a macroscopic system that is forced into a superposition by some strong external force. Instead of a gentle and continuous spreading of the wavepacket (such as the one caused by the environment, which is subject to the instability discussed before), consider a quantum mechanical operation which quickly drives a macroscopic system into a superposition of ordered states with well separated orientations of their order parameters (the instantaneous coupling of the order parameter to a quantum superposition would in general do the trick). To be specific, consider the initial state 
\begin{align}
\left| \psi(0) \right> = \alpha \left| AFM \right>_x + \beta \left| AFM \right>_z.
\label{psi0}
\end{align}
Here $|AFM\rangle_x$ signifies an antiferromagnetic N\'eel state ordered along the $x$ axis. The time evolution of the order parameter measured along the $z$ axis, starting from the initial state with $\alpha=\beta=\sqrt{1/2}$ is shown in figure \ref{theta}. Here we again consider a constant symmetry breaking field $b$ and the time evolution defined by equation~\eqref{psidot}.

\begin{figure}[t]
\center
\includegraphics[width=0.8\columnwidth]{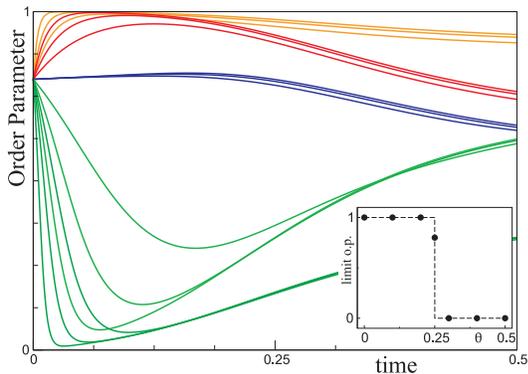}
\caption{(Color online) The evolution of the order parameter as a function of time (in units of $\hbar s$) starting from the superposition state $\sqrt{ {1/5}} \left| AFM \right>_x + \sqrt{ {4/5}} \left| AFM \right>_z$. Each set of three curves represents the time evolution with three different values for $N$ in the presence of a single, constant orientation of the unitarity breaking field $b$. From top to bottom the angle between $b$ and the $z$ axis for the different sets is $0.1~\pi$, $0.2~\pi$, $0.25~\pi$, $0.3~\pi$ and $0.4~\pi$. The point at which the initial fast reduction process starts favouring the $x$ orientation over the $z$ orientation is seen to be at $0.25~\pi$.}
\label{bornplot}
\end{figure}

The evolution of this initial state can be seen as a a combination of the two processes encountered before. First there is a fast reduction of the initial state to a single ordered state within a timescale $\propto 1/(Nb)$. The choice of which ordered state results from this fast initial evolution depends only on the chosen direction of the unitarity breaking field, and not on the weights of the different ordered states in $|\psi(0)\rangle$ (as can be seen in figure~\ref{bornplot}). After the fast reduction to a single ordered state, the slow process of rotating the order parameter towards alignment with the field $b$ takes over. This secondary process happens in a time which scales as $\propto \sqrt{1/(Jb)}$. In the limit that the number of particles goes to infinity before the unitarity breaking field is sent to zero, the result is thus a spontaneous, instantaneous reduction of the initial state to just a single one of the ordered states present in the original superposition.

The observation that the selection of the ordered state to be singled out by the spontaneous dynamics depends on the chosen (constant) orientation of $b$ signifies the fact that the initial state is unstable with respect to two different and competing perturbations: one for each orientation of the order parameter present in the initial superposition. The two possible stable final states are mutually exclusive since for any choice of unitarity breaking field, only one orientation of the staggered magnetisation results. 

As mentioned before, the dynamical nature of the symmetry breaking process implies that we should really consider a time-dependent, fluctuating symmetry breaking field rather than only a constant field. In the presence of such a fluctuating field, it is clear that there must be a competition between the two instabilities of the initial state. In general, this gives rise to a statistical outcome of the reduction process (just like the instabilities of the singlet state gave rise to a statistical, random selection of the orientation of its order parameter under equilibrium conditions). The resulting dynamic process could be somewhat reminiscent of the evolutions considered in the GRW and CSL models for quantum state reduction~\cite{Pearle89,Ghirardi90}, and consist of a random sequence of amplifying one or the other ordered state until one of them completely dominates. 

It was shown recently by Zurek, using the concept of ENVariance~\cite{Zurek03}, that one can obtain conclusions about the statistics of the final results of a dynamic competition between instabilities such as the one considered here, without knowing the exact dynamics governing the competition process~\cite{Zurek05}. It is shown in the appendix that Zurek's proof is applicable here without the need for any assumptions regarding our system. Following his derivation one finds that the only possible result of the dynamic competition between different instabilities of the initial state of equation~\eqref{psi0} is the emergence of Born's rule: the probability of a certain direction of the order parameter emerging from the process is given by the square of its weight in the initial wavefunction~\cite{Born26}. Notice that this result is not an expectation value: it is valid even for the quantum dynamics of a single macroscopic object that is forced into a superposition state.

\subsection{6: Experimental Predictions} \noindent
The dynamic spontaneous symmetry breaking process described in the previous sections results in unaltered, purely unitary quantum dynamics for microscopic particles, but also gives rise to spontaneous and non-unitary effects in the thermodynamic limit. For truly macroscopic objects the non-unitarity will be effectively instantaneous, and the quantum dynamics of such objects correspondingly reduces to classical physics. Somewhere in between the micro and macro scales however, there must be a class of mesoscopic objects which are just sensitive enough to the presence of a small (but finite) time translation symmetry breaking field to undergo non-unitary dynamics on timescales that are measurable by human standards. The scale at which this happens should in fact be the same scale at which collections of interacting quantum particles become large enough to be meaningfully ascribed a (stable) orderparameter and considered classical, symmetry broken objects under equilibrium conditions. This prediction can in principle be exploited to experimentally test the ideas which are put forward in this paper.

The greatest obstacle in realising such an experimental test will be decoherence. Quite apart from the issue of its applicability to only ensemble averages, decoherence is of course a real physical phenomenon which severely complicates the observation of quantum effects in systems coupled to a reservoir. To observe the breakdown of unitary quantum dynamics, one will thus have to find a way to experimentally distinguish its effects from those of the usual environmental decoherence. The most obvious way of doing that is to look at single-shot experiments only. If the famous experiment of Zeilinger et al.~\cite{Zeilinger99}, interfering C$_{60}$ molecules one at a time, could be scaled up to truly macroscopic proportions, it would form the ideal testing ground for observing the transition from quantum to classical behaviour. The crossover scale could then be directly compared with the scale at which ordering and rigidity appear under equilibrium conditions, and this could be used to examine the role played by dynamic spontaneous symmetry breaking. Such macroscopic interference experiments however, seem to be very far from what can presently be experimentally realised.

We thus have to look for a different Schr\"odinger-cat like state of a mesoscopic system which is large enough to feel the effects of non-unitarity, but small enough to still have a measurably long reduction time. Creating such mesoscopic superpositions in the lab surely is not an easy task, but significant experimental progress towards its realisation is already being made in setups in for example quantum computation (superconducting flux qubits and Cooper pair boxes) or cold atom physics (Bose Einstein condensates in optical traps). Note however that the superposition must be a combination of states with different orientations of the order parameter itself. Superpositions of elementary excitations (such as magnons, phonons or supercurrents) which do not affect the order parameter, are not subject to the spontaneous reduction process described in the previous sections, even in a truly infinite system. Although distinguishing any non-unitary dynamics from the effects of decoherence is known to be very hard in most systems~\cite{vanWezel:penrose}, there is at least one experimental arena in which there seems to be, at least in principle, an opportunity for doing so: the Bose-Einstein condensation of exciton polaritons in semiconductor microcavities~\cite{Kasprzak06,Keeling07,Wouters07}.

Exciton polaritons are composite particles built partly from particle-hole pairs (excitons) and partly from photons. This unique combination of light and matter allows the particles to have strong interactions (due to their excitonic nature) while also being susceptible to direct experimental manipulation (due to their coupling to light). Although the short lifetime of the excitons implies that the condensate formed from polaritons in semiconductor microcavities is necessarily in a dynamical rather than a thermal equilibrium, it has been shown that the condensed phase shares many properties of the usual atomic Bose-Einstein condensate: it is a coherent state of spontaneously broken symmetry with an associated Goldstone mode~\cite{Wouters07}. Recently it has been proposed that the dynamical nature of the polariton condensation can be used to explicitly break the U(1) phase symmetry present in a continuously, resonantly pumped experiment using an additional continuous probing laser~\cite{Wouters07,Amo07}. The coherence of the condensate can be independently tested by looking at the coherence and polarisation of the light emitted by recombining excitons~\cite{Ciuti01}. If the pumping power is large enough to create a polariton condensate in a truly classical, symmetry broken state, then the condensate should retain its coherence even after the probing laser has been turned off. At lower power the condensate wavefunction will instead spread out over phase space and look symmetric again. Building on these results, the following experiment comes to mind. One can use the lack of number conservation in the condensate's dynamical equilibrium~\cite{Amo07}, to create a superposition of different order parameters by subjecting the polaritons to a superposition of two different probing laser beams. The resulting macroscopic superposition is then expected to spontaneously collapse into just one ordered state for high enough pumping power due to dynamical spontaneous symmetry breaking, while lower pumping power (and the absence of symmetry breaking) should lead only to quantum beatings between the states of the initial superposition. If the transition from collapse behaviour to quantum beatings occurs at the same pumping power at which a single condensate has been seen to remain stable after turning off the probing laser, that would form a strong experimental indication of the involvement of dynamic spontaneous symmetry breaking.

\subsection{7: Conclusions} \noindent
In summary, we have shown here that macroscopic objects which spontaneously break a continuous symmetry under equilibrium conditions are also subject to a spontaneous breakdown of quantum mechanics' unitary time translation symmetry. The coincidence of objects liable to dynamic spontaneous symmetry breaking with those liable to equilibrium spontaneous symmetry breaking is ensured by the crucial role played by the thin spectrum which is known to characterise the latter objects. Dynamic spontaneous symmetry breaking augments the well known theories of equilibrium spontaneous symmetry breaking and decoherence in the domains where these theories do not apply, and so leads to the symmetry broken state being not just the only stable ground state under equilibrium conditions, but also the only stable state dynamically. The quantum dynamics of any symmetric state, and more generally any superposition of differently ordered states, is almost infinitely sensitive to non-unitary perturbations in the thermodynamic limit, and such states must thus spontaneously and instantaneously be reduced to a state with only a single order parameter.

Applying this description of dynamic spontaneous symmetry breaking to the ordered states in our classical world, it becomes clear why these ordered classical states do not seem to be bothered by the interaction with their quantum environments: any buildup of quantum uncertainty is immediately reduced by the dynamical symmetry breaking process. Using the description instead to study the fate of a superposition of different classical states, one finds that only a single state can survive the spontaneous breakdown of quantum dynamics, and that the probability for finding any one particular outcome must be given by Born's rule.

The predicted spontaneous breaking of unitary quantum time evolution can in principle be tested experimentally if one has a controlled way of constructing superpositions of differently ordered mesoscopic states. One type of system in which this may possibly be achieved is given by the polariton condensates in which the phase of the order parameter can be selected using the coherence of an incident laser beam.

\subsection{Acknowledgements} \noindent
I would like to gratefully acknowledge countless stimulating and insightful discussions with Jeroen van den Brink and Jan Zaanen.

\subsection{Appendix A: Quantum Measurement} \noindent
In the main text we investigated the stability of a macroscopic state created by a quantum mechanical operation which quickly drives an ordered system into a superposition of differently ordered states with well separated orientations of their order parameters. 
One instance in which such a process is believed to occur is quantum measurement. By its very nature a quantum measurement is defined to be a process in which some property of a microscopic quantum state is translated into a specific pointer state of a macroscopic measurement machine~\cite{Zurek81,Joos85}. The different pointer states of such a machine must be easily distinguishable, classical states. In practise this always implies that they are symmetry broken states with different values or orientations for their order parameters. If we take these properties of the measurement machine at face value then it is clear that the measurement of a superposed quantum state must also lead to a superposition of pointer states in the measurement device because of the unitarity of quantum mechanical time evolution~\cite{Bassi}. This simple observation already lead John von Neumann to postulate a collapse process which takes place after the usual quantum mechanical time evolution, and acts only on macroscopic superpositions~\cite{Neumann55}. The explanation of why the collapse process exists, why it only acts on pointer states and not on microscopic states and why it gives rise to Born's rule (dictating the probability of a certain outcome) is known as the quantum measurement problem. Many attempts have been made to either introduce a specific collapse process into quantum mechanics or to avoid the problem altogether by interpreting the mathematics of quantum mechanics in a different way. However, neither of these approaches has yet lead to a satisfactory resolution of all of the questions posed by the measurement problem.

Our analysis of the quantum dynamics of a superposition of differently ordered states in the thermodynamic limit suggests the following description of quantum measurement: a quantum measurement machine is any system with a well developed order parameter that can be coupled to a microscopic quantum system in such a way that the orientation of the order parameter after the coupling process has been completed, represents the property of the microscopic state that is to be measured. In general such a coupling should give rise to macroscopic superpositions of the order parameter, but the dynamical, spontaneous breakdown of quantum mechanics' unitary time evolution ensures the spontaneous reduction of such superpositions into a state with just a single well defined order parameter. Because the macroscopic superposition state is subject to multiple competing instabilities, the outcome of the reduction process is probabilistic. The probability for obtaining a specific outcome is automatically guaranteed to agree with Born's rule due to the properties of the process of dynamic spontaneous symmetry breaking (see also Appendix B).

Using dynamic spontaneous symmetry breaking, we have arrived at a clear-cut definition of what a measurement machine is; why it is subject to a collapse process; why this collapse does not influence microscopic quantum states; and we have recovered Born's rule. The quantum measurement problem is thus reduced to the problem of identifying possible sources of non-unitary perturbations to the theory of quantum mechanics, which could drive the dynamic spontaneous symmetry breaking process. Regardless of its source, any non-unitary influence which can couple to a suitable order parameter will be amplified by the symmetry breaking process, and yield the expected macroscopic dynamics.
%

\subsection{Appendix B: Detailed Derivation of Born's Rule} \noindent
In this appendix we will give the detailed derivation of the emergence of Born's rule from the dynamic spontaneous breaking of quantum mechanical time translation symmetry as applied to the case of the Lieb-Mattis antiferromagnet. There are three main requirements that need to be satisfied in order for the following derivation to be applicable. These requirements are: (1) The spontaneous evolution must yield a final state with only a single orientation of the order parameter, and the selection of the specific order parameter to be realised must be a probabilistic process; (2) The probability of obtaining a certain outcome may only depend on its weight in the initial superposition; (3) If the initial superposed state is entangled with some other, external quantum mechanical object with which the antiferromagnet has no further interaction, then the probability for finding a certain final orientation of the antiferromagnetic order parameter should not be affected by the precise state of the external quantum mechanical object.

To see that these requirements are all satisfied by the process of dynamic spontaneous symmetry breaking described before, consider the initial state
\begin{align}
\left| \psi(0) \right> = \alpha \left| e1 \right> \otimes \left| AFM \right>_x + \beta \left| e2 \right> \otimes \left| AFM \right>_z,
\label{alphabeta}
\end{align}
where $|\alpha|^2 + |\beta|^2=1$, $|AFM\rangle_x$ is the state with full antiferromagnetic order along the $x$ axis, and the states $|e1\rangle$ and $|e2\rangle$ are some external states which have no further interaction with the antiferromagnet whatsoever. The Hilbert space of the combined system of antiferromagnet and external states can be written as a product of the space of states of the antiferromagnet and the space of external states. Following the discussion of the quantum dynamics of a superposed macroscopic state in the main text, it is clear that the dynamics of the initial state $\left| \psi(0) \right>$ is unstable with respect to two orientations of the symmetry breaking field. Since the two instabilities of $\left| \psi(0) \right>$ must compete with each other, only one of the two available stable states can be realised, and the selection of which state is realised in the presence of a fluctuating symmetry breaking field is a probabilistic process, as stated in requirement one. Furthermore, since the competition between instabilities takes place on an infinitesimally short timescale, it cannot be influenced by the finite energy scale $J$. The {\it fluctuating} field $b$ is guaranteed by symmetry not to favour either one of the two possible final states. The only thing left to determine the probability of finding a certain final state is then the choice of the initial state itself: i.e. only the weights $\alpha$ and $\beta$ can determine the probability distribution of final states, in agreement with requirement two. That these weights in fact do influence the probability distribution is obvious from the fact that the initial states with $\alpha$ or $\beta$ equal to zero are stable states. The external states $|e1\rangle$ and $|e2\rangle$ cannot influence the spontaneous dynamics because all of the competition between the instabilities is governed by the unitarity breaking field $b$. This field acts only on the states of the antiferromagnet, and not on any other part of the Hilbert space (requirement three). The initial state $\left| \psi(0) \right>$ will thus be spontaneously and instantaneously reduced to either the state $\left| e1 \right> \otimes \left| AFM \right>_x$ or the state $\left| e2 \right> \otimes \left| AFM \right>_z$, while the probabilities $P_x(\psi)$ and $P_z(\psi)$ for finding either final state depend only on the values of $\alpha$ and $\beta$.

Building on these known properties of the final probabilities, let's now follow Zurek's arguments for obtaining the exact final probability distribution~\cite{Zurek05}. First consider two different initial states:
\begin{align}
\left| \psi \right> = \alpha \left| e1 \right> \otimes \left| AFM \right>_x + \beta \left| e2 \right> \otimes \left| AFM \right>_z \phantom{.} \nonumber \\
\left| \phi \right> = \alpha \left| e3 \right> \otimes \left| AFM \right>_x + \beta \left| e4 \right> \otimes \left| AFM \right>_z.
\label{statementA}
\end{align}
Since the final probabilities can only depend on the weights of the classical states in the initial wavefunction (req. 2), it is immediately clear that $P_x(\psi)=P_x(\phi)$. This must hold independent of the external states $|e1\rangle$ through $|e4\rangle$ (req. 3), and thus it must also hold in the special case $|e1\rangle=e^{i \theta} |e3\rangle,~ |e2\rangle=|e4\rangle$, showing that the probability distribution cannot depend on the phases of the weights in the initial wavefunction.

Next, consider the initial states
\begin{align}
\left| \psi \right> = \alpha \left| e1 \right> \otimes \left| AFM \right>_x + \beta \left| e2 \right> \otimes \left| AFM \right>_z \phantom{.} \nonumber \\
\left| \chi \right> = \alpha \left| e2 \right> \otimes \left| AFM \right>_z + \beta \left| e1 \right> \otimes \left| AFM \right>_x.
\label{statementB}
\end{align}
Clearly, we must have $P_x(\psi)=P_z(\chi)$ for any choice of $\alpha$ and $\beta$. In the special case $|\alpha|=|\beta|$ we also know $P_z(\psi)=P_z(\chi)$, and thus we find that in that case $P_x(\psi)=P_z(\psi)$. In other words, if the sizes of the weights corresponding to two final states are equal, then so are the probabilities for finding these states. This statement can be trivially extended to yield the rule that a set of possible final states with equal weights in the initial wavefunction leads to equal probability for finding any one of the final states within that set. Continuing that line of thought, consider
\begin{align}
\left| \psi \right> = \alpha \left| AFM \right>_i + \alpha \left| AFM \right>_j+ \alpha \left| AFM \right>_k + ...
\label{statementD}
\end{align}
where $i$, $j$ and $k$ are different directions in real space. The combined probability $P_{i~\text{or}~j}(\psi)$ must then be equal to $P_i(\psi)+P_j(\psi)=2P_k(\psi)$, which follows directly from the additivity of probabilities and the mutual exclusivity of the three possible final states. That the final states are in fact mutually exclusive is guaranteed by requirement 1: in the thermodynamic limit $\left| AFM \right>_i$ and $\left| AFM \right>_j$ correspond to states with different directions of their order parameters, which can have no overlap and only one of which can be the result of the spontaneous dynamics. Extending this result, it is now clear that within a set of possible final states with equal weights in the initial wavefunction, a subset has a combined probability equal to the relative size of the subset times the total probability of the entire set.

Finally, consider the initial state
\begin{align}
\hspace{-5pt} \left| \psi \right> = \sqrt{\frac{m}{N}} \left| e1 \right> \otimes \left| AFM \right>_x + \sqrt{\frac{n}{N}} \left| e2 \right> \otimes \left| AFM \right>_z.
\label{statementE}
\end{align}
The probability $P_x(\psi)$ is independent of the external states (req. 3). We are therefore free to write $|e1\rangle$ and $|e2\rangle$ in a basis in which they are a sum of states with equal weights (such a basis can be shown to always exist~\cite{Zurek05}):\begin{align}
\left| e1 \right> &= \sqrt{\frac{1}{m}} \left[ \left| E1_1 \right> + \left| E1_2 \right> + ... + \left| E1_m \right> \right] \phantom{.} \nonumber \\
\left| e2 \right> &= \sqrt{\frac{1}{n}} \left[ \left| E2_1 \right> + \left| E2_2 \right> + ... + \left| E2_n \right> \right] .
\label{statementE2}
\end{align}
Reinserting these definitions into equation \eqref{statementE} yields
\begin{align}
\left| \psi \right> = \sqrt{\frac{1}{N}} \left[  \sum_{i=1}^m \left| E1_i \right>  \otimes \left| AFM \right>_x + \right. \nonumber \\
\left. \sum_{j=1}^n \left| E2_j \right> \otimes \left| AFM \right>_z \right] .
\label{statementE3}
\end{align}
In this expression all weights are equal, and using the previously found rules we must thus conclude that $P_x(\psi)=\frac{n}{m}P_z(\psi)$. In the case that the total probability for finding any outcome at all is one, this result precisely corresponds to Born's rule: the probability for finding any specific final orientation of the order parameter is equal to the square of the weight of the corresponding state in the initial wavefunction~\cite{Born26}. The extension of this result to include also weights which are square roots of non-rational numbers is trivial because the rational numbers are dense on the real line~\cite{Zurek05}.


\end{document}